\documentclass[reprint,amsmath,amssymb,aps,preprintnumbers]{revtex4-2}

\usepackage{natbib,xcolor}

\usepackage{soul}

\UseRawInputEncoding 

\begin{document}

\title{
On black hole supertranslations and hydrodynamic enstrophy
}

\preprint{MIT-CTP/5340}

\author{Raja Marjieh}
\email{rajamarjieh@gmail.com}
\affiliation{Department of Psychology, Princeton University, Princeton, NJ 08544, USA}
\author{Natalia Pinzani-Fokeeva}
\email{pinzani@mit.edu}
\affiliation{Center for Theoretical Physics, Massachusetts Institute of Technology, Cambridge, MA 02139, USA, and\\
Department of Physics, University of Florence,  Via G. Sansone 1, I-50019, Sesto Fiorentino, Firenze, Italy}
\author{Bar Tavor}
\email{btavor@campus.technion.ac.il}
\author{Amos Yarom}
\email{ayarom@physics.technion.ac.il}
\affiliation{Department of Physics, Technion, Haifa 32000, Israel}

\date{\today}
\begin{abstract}
We study the relation between approximate horizon symmetries of AdS black branes and approximately conserved currents in their dual hydrodynamic description. 
We argue that the existence of an approximately conserved enstrophy current unique to $2+1$ dimensional fluid flow implies that AdS${}_4$ black branes  possess a special class of approximate supertranslations (which we identify).
\end{abstract}
\maketitle


\section{Introduction}
\label{S:intro}

Enstrophy is a scalar quantity associated with non relativistic, incompressible, fluid flow. In two spatial dimensions it can not increase in time, and in the absence of dissipation it is conserved.  
Its properties are key ingredients in generating the unique features of turbulent flows in two spatial dimensions. 

The study of enstrophy in relativistic fluids 
is lagging behind that of its Galilean cousin.
 It is possible to identify a relativistic enstrophy current which is conserved for an ideal fluid in $2+1$ dimensions  \cite{Carrasco:2012nf,Marjieh:2020odp,Pinzani-Fokeeva:2021klb}.
Since dissipative effects spoil this property,  we will refer to such a current as being approximately conserved.

The existence of a relativistic enstrophy current suggests, via the gauge-gravity duality \cite{Bhattacharyya:2008jc}, that AdS${}_4$ black branes possess an approximate symmetry associated with approximate enstrophy conservation.
Since the enstrophy current exists only in the hydrodynamic limit, one might expect that an associated approximate black hole symmetry will exist on the event horizon.

Indeed, it has recently been established that stationary black hole geometries are endowed with horizon symmetries, classified as supertranslations and superrotations \cite{Donnay:2015abr,Donnay:2016ejv} (see \cite{Carlip:1999cy,Hotta:2000gx,Koga:2001vq,Hotta:2002mq} for older work on this topic). 
Generic fluid flows which possess enstrophy are not stationary and imply non stationary dual black hole configurations. While the definition of supertranslations and superrotations may be extended to non stationary black holes, they are generally not associated with conserved currents \cite{Chandrasekaran:2018aop}.

In this work we will  relate a subset of supertranslation transformations of AdS$_{4}$ black branes 
to approximate enstrophy current conservation in the dual hydrodynamic  description of the field theory. 
Further, we clarify the role of the remaining generators of supertranslations and identify them with symmetries leading to non local approximately conserved  currents. Superrotations do not lead to approximately conserved currents in a dual hydrodynamic setting.

 Relations between hydrodynamics and horizon symmetries have been considered  in the past. In \cite{Eling:2016xlx}, the authors analysed  asymptotically AdS black brane solutions dual to a superfluid flow and identified the action of supertranslations on horizon data with the superfluid Goldstone mode. In \cite{Penna:2017bdn}, horizon symmetries in $3+1$ dimensions were shown to be in one to one correspondence with symmetries of compressible non relativistic fluids in $2+1$ dimensions (see also \cite{Donnelly:2020xgu} for a similar relation in a different context). The authors of \cite{Donnay:2019jiz} related the dynamics of the horizon to Carrollian fluids and the horizon symmetries to Carrollian geometry. While the authors did not discuss this, their results are suggestive of a Carrollian enstrophy current [3] associated with horizon symmetries. Be that as it may, the novelty of the current work is in its explicit identification of (a subset of) horizon symmetries of asymptotically AdS black holes with the symmetries of relativistic fluid dynamics of the boundary theory.

Our exposition starts with a review of the relativistic, conformal, enstrophy current, as first discussed in \cite{Carrasco:2012nf}, and its extensions \cite{Marjieh:2020odp,Pinzani-Fokeeva:2021klb}.  
Using the construction of \cite{Chandrasekaran:2018aop}, we then discuss the general structure of horizon preserving diffeomorphisms.
Finally, we use our knowledge of enstrophy conservation in the boundary theory to show that AdS${}_4$ black brane geometries possess special supertranslation symmetries. We end with a discussion and outlook.

\section{The enstrophy current}
The equations of motion of an uncharged, conformal, relativistic fluid are given by
\begin{equation}
\label{E:TEOM}
	\nabla_{\mu}T^{\mu\nu} = 0\,,
\end{equation}
where $T^{\mu\nu}$ is the energy momentum tensor of the fluid and is a local function of the the fluid velocity $u^{\mu}$ (satisfying $u^{\mu}u_{\mu}=-1$) and fluid temperature $T$. Working in a derivative expansion we have
\begin{equation}
\label{E:constitutive}
	T^{\mu\nu} = P(T) \left( (d+1) \,u^{\mu}u^{\nu} + g^{\mu\nu} \right)   + \mathcal{O}(\nabla)
\end{equation}
where $d>1$ is the number of spatial dimensions, $P(T) = p_0 T^{d+1}$ is the thermodynamic pressure with $p_0$ a positive real number,
and $\mathcal{O}(\nabla)$ denotes expressions which contain one or more derivatives of the hydrodynamic variables $u^{\mu}$ and $T$. 

For any function $g$ which satisfies 
\begin{equation}
\label{E:fconstraint}
	u^{\mu} \nabla_{\mu} g = {\cal O}(\nabla^{[g]+2})\,,
\end{equation}
we may construct the current
\begin{equation}
\label{E:Jgdef}
	J_g^{\mu} = g\, p_0 (d+1)\, T^{d} u^{\mu}
\end{equation}
which satisfies
\begin{equation}
\label{E:divJf}
	\nabla_{\mu} J_g^{\mu} = \mathcal{O}(\nabla^{[g]+2})\,.
\end{equation}
In obtaining \eqref{E:divJf} we have used that $\nabla_{\mu} (T^{d} u^{\mu} )= \mathcal{O}(\nabla^2)$ as a result of the equations of motion \eqref{E:TEOM} expanded in derivatives. The particular choice of the overall constant $p_0 (d+1)$ will become clear shortly. 

A naive power counting argument would suggest that $\nabla_{\mu}J_g^{\mu}$ is of the same order as $g$ plus one in a derivative expansion. Instead, \eqref{E:divJf} implies that it is order $g$ plus two. In other words, it is conserved at least to leading order in a derivative expansion. In what follows we will refer to $J_g^{\mu}$ as an approximately conserved current.

We often require conserved currents to be local in the hydrodynamic variables, in this case $u^{\mu}$ and $T$.
While there are many non local solutions to \eqref{E:fconstraint}, local solutions are more difficult to come by. Clearly, $g=1$, or any constant for that matter, is a solution to \eqref{E:fconstraint}. For such solutions we obtain
\begin{equation}
\label{eq:entropy}
	J_1^{\mu} = s u^{\mu}
\end{equation}
with $s = \frac{\partial P}{\partial T}$ the entropy density. Approximate conservation of $J_1^{\mu}$ coincides with the leading order equation of motion and implies conservation of entropy in the absence of dissipation.

Another solution to \eqref{E:fconstraint} which is local in the hydrodynamic variables is given by
\begin{equation}
	g = \frac{\Omega_{\alpha\beta}\Omega^{\alpha\beta}}{s^2}\,,
\end{equation}
with
\begin{equation}
	\Omega_{\mu\nu} = \partial_{\mu} \left( T u_{\nu} \right) - \partial_{\nu} \left( T u_{\mu} \right)\,,
\end{equation}
and is valid only in $2+1$ spacetime dimensions. The associated current
\begin{equation}
\label{E:Jgeneral}
	J_{\frac{\Omega^2}{s^2}}^{\mu} = \frac{\Omega_{\alpha\beta}\Omega^{\alpha\beta}}{s} u^{\mu}
\end{equation}
is the relativistic enstrophy current \cite{Carrasco:2012nf,Marjieh:2020odp,Pinzani-Fokeeva:2021klb}, and its associated charge is referred to as enstrophy. Of course, if $g$ solves \eqref{E:fconstraint} so do powers of $g$, and we find a set of conserved currents,
\begin{equation}\label{E:generaln}
	J_{\left(\frac{\Omega^2}{s^2}\right)^n}^{\mu} = \left(\frac{\Omega_{\alpha\beta}\Omega^{\alpha\beta}}{s^2}\right)^n s u^{\mu}\,.
\end{equation}	
When   $n=0$, the above expression reduces to the entropy current \eqref{eq:entropy}.  The currents \eqref{eq:entropy} and \eqref{E:generaln} comprise the only known local solutions to \eqref{E:fconstraint}.

We note in passing that in a non relativistic, incompressible fluid the charges
	$\int \left(\omega_{ij}\omega^{ij} \right)^n d^2x$,
with $\omega_{ij} = \partial_i v_j - \partial_j v_i$ and $v_i$ the velocity field, are conserved in the inviscid limit. 
The former integral with $n=1$ is referred to as the total enstrophy.
Once  dissipative effects are included, the total enstrophy decreases in time. This property, together with energy conservation, leads to an inverse energy cascade in 2+1 dimensional incompressible  non relativistic turbulent flows whereby energy is transferred from small to  large scales \cite{Kraichnan}. Whether similar statements can be made for relativistic fluids is yet an open problem. 

\section{Horizon symmetries and charges}

An extensive analysis of horizon symmetries and charges was carried out in \cite{Chandrasekaran:2018aop}. In what follows we summarize the essential ingredients of \cite{Chandrasekaran:2018aop} required for this work. Consider a spacetime $\mathcal{M}$ with metric $g_{ab}$ and event horizon $\mathcal{N}$ whose topology is $\mathcal{Z} \times \mathbb{R}$ with $\mathbb{R}$ a null direction. 
We refer to $\mathcal{Z}$ as the base space of $\mathcal{N}$. In the case of an asymptotically AdS${}_{d+2}$ black brane, $\mathcal{Z}$ has topology $\mathbb{R}^{d}$. Let us denote the pullback to the horizon by $\Pi_i{}^a$ so that $g_{ij} = \Pi_j{}^b \Pi_j{}^b g_{ab}$ is the induced metric on the event horizon. Since $\mathcal{N}$ is null, $g_{ij}$ is not invertible.  We denote the pullback from $\mathcal{N}$ to the base space $\mathcal{Z}$ by $\Pi_A{}^i$ so that the induced metric on $\mathcal{Z}$ is $g_{AB}=\Pi_A{}^a \Pi_B{}^b g_{ab}$ where $\Pi_A{}^a = \Pi_A{}^i \Pi_i{}^a$. Since $\mathcal{Z}$ is spacelike, $g_{AB}$ is invertible. In what follows we will consistently use $a,\,b,\,\ldots$ for indices on $\mathcal{M}$, $i,\,j,\,\ldots$ for indices on $\mathcal{N}$ and $A,\,B,\,\ldots$ for indices on $\mathcal{Z}$. Later, when we will focus on asymptotically AdS spaces, we will introduce indices  $\mu,\,\nu,\ldots$ on the asymptotic boundary of the spacetime.

We denote by $\ell^i$ a representative vector field generating null geodesics along the null direction $\mathbb{R}$ of $\mathcal{N}$, and by $\ell^a$ an extension of it to all of $\mathcal{M}$. We define the  non affinity parameter $\kappa$ via
\begin{equation}
\label{E:defkappa}
	\ell^a \nabla_a \ell^b\Big|_{\mathcal{N}} = \kappa \ell^b\Big|_{\mathcal{N}}\,,
\end{equation}
where, as usual, $\nabla_a$ is the covariant derivative on ${\mathcal{M}}$.
We also denote  a null cotangent vector on $\mathcal{N}$ by $n_i$ and normalize it and its extension to $\mathcal{M}$ such that $n_a \ell^a=-1$. 

We will often need to go back and forth between tangent vectors on $\mathcal{M}$ (or $\mathcal{N}$) and tangent vectors on $\mathcal{N}$ (or $\mathcal{Z}$). For instance, suppose that $v^a \in T(\mathcal{M})$ satisfies $v^a \ell_a\Big|_{\mathcal{N} }= 0$ (recall that  $\ell_a$ is the normal to $\mathcal{N}$). Then, we may always define a unique $v^i \in T(\mathcal{N})$ such that $v^a w_a \Big|_{\mathcal{N}} = v^i \Pi_i{}^a w_a$ for any $w_a \in T^*(\mathcal{M})$. To simplify our notation we will write \begin{equation}
	v^a \partial_a \cong v^i \partial_i\,.
\end{equation}

An infinitesimal coordinate transformation $\chi = \chi^i \partial_i$ is referred to as a generator of a horizon preserving diffeomorphism if 
\begin{equation}
\label{E:preserving}
	\pounds_{\chi} \ell^{i} = {\beta}\ell^{i} \,,
	\qquad
	\pounds_{\chi} \kappa  =  (\beta \kappa+\pounds_{\ell}\beta),
\end{equation}
with $\beta$ a function on $\mathcal{N}$. The first equation in \eqref{E:preserving} is an infinitesimal version of a rescaling of the null vector $\ell^i\partial_i$ (which does not have a well defined length due to the fact that it's null). The second equation in \eqref{E:preserving} corresponds to a shift in the non affinity parameter resulting from a rescaling of $\ell^a \partial_a$, c.f., \eqref{E:defkappa}.

The generator $\chi^i$ can be naturally decomposed into a component parallel to $\ell^i$ and a component orthogonal to it, 
\begin{equation}
\label{E:xidecomposition}
	\chi^i \partial_i = X^i\partial_i  + f \ell^i \partial_i\,,
\end{equation}
with $X^i n_i=0$. With this decomposition equation \eqref{E:preserving} reads
\begin{align}
\begin{split}
\label{E:supers}
	\pounds_{\ell} X^i \propto\, & \ell^i\,,\\
	\qquad
	\pounds_\ell \left(\pounds_\ell  + \kappa \right) f +
	X^i \pounds_{\ell} \left(\pounds_{\ell} + \kappa \right) n_i+ \pounds_X \kappa = \,& 0\,.
\end{split}
\end{align}
It is tempting to refer to a horizon symmetry associated with $X^i$ as a superrotation and to a horizon symmetry associated with $f\ell^i$ as a supertranslation. Note, however, that the distinction between the two is dependent on the choice of $n_i$. In \cite{Chandrasekaran:2018aop} it was shown, using an explicit construction, that there exists an $n_i$ for which
\begin{equation}
\label{E:n}
	\pounds_{\ell} \left(\pounds_{\ell} + \kappa \right) n_i +\partial_i \kappa=0\,.
\end{equation}
With this choice of $n_i$, equations \eqref{E:supers} can be shown to reduce to
\begin{equation}
\label{E:canonicaln}
	\pounds_{\ell} \left(\pounds_{\ell}+\kappa\right)f= 0\,. 
\end{equation}
One may refer to the supertranslations and superrotations obtained using the $n_i$ which leads to \eqref{E:canonicaln} as  canonical supertranslations and superrotations.

One of the results of \cite{Chandrasekaran:2018aop} is that, in the absence of matter, we can associate to each such $\chi$ a Wald-Zoupas charge, $Q_{\chi}$, which is conserved whenever the horizon is stationary. Operatively, for each generator $\chi$ we define a current $q_{\chi}^{i}$ satisfying
\begin{equation}
\label{E:qdef}
	q_{\chi}^{j} = \left(\chi^i \mathcal{K}_i{}^j - \theta \chi^j - \beta \ell^j \right)\,.
\end{equation}
Here $\beta$ is 
associated with the scaling of $\ell^i$ under $\chi^j$ as in \eqref{E:preserving}, $\mathcal{K}_i{}^j$ denotes the Weingarten map
\begin{equation}
	\mathcal{K}_i{}^j \partial_j \cong \Pi_i{}^a \nabla_a \ell^b \partial_b \,,
\end{equation}
and $\theta$ is the expansion associated with $\ell^i$,  
\begin{equation}\label{E:horder}
	\theta = \nabla_i \ell^i = \frac{1}{\sqrt{|g_{AB}|} }\partial_i \left( \sqrt{|g_{AB}|} \ell^i \right)\,.
\end{equation}

It is straightforward though somewhat tedious to compute the divergence of $q_{\chi}^{i}$.
Recall that the Weingarten map satisfies
\begin{equation}
\label{E:Weingartenproperties}
	\ell^i \mathcal{K}_i{}^j = \kappa \ell^j\,,
	\qquad
	\mathcal{K}_i{}^j g_{jk} = \frac{1}{2} \pounds_{\ell} g_{ik}\,.
\end{equation}
The expression on the right hand-side of the second equality is the second fundamental form on $\mathcal{N}$, $K_{ij} = \frac{1}{2} \pounds_{\ell} g_{ij}$.
 It is orthogonal to $\ell^i$ allowing us to write
\begin{equation}
	K_{ij} = \frac{\theta}{d}g_{ij} + \Sigma_{ij}
\end{equation}
where $\Sigma_{ij}dx^i dx^j  \cong  \Sigma_{AB} dx^A dx^B$ is symmetric and traceless, $\Sigma_{AB} g^{AB}=0$. Thus, the most general expression for $\mathcal{K}_i{}^j$ satisfying \eqref{E:Weingartenproperties} is
\begin{equation}
\label{E:Kappadecomposition}
	\mathcal{K}_i{}^j =  \omega_i \ell^j   +S_i{}^j\ \,,
\end{equation}
where 
\begin{equation}
	S_i{}^j dx^i \partial_j \cong \left(\frac{\theta}{d} \delta_A{}^B + \Sigma_A{}^B\right) dx^A \partial_B\,,
\end{equation}
and
\begin{equation}
	\omega_i =  -\kappa n_i dx^i + \Omega_i dx^i 
\end{equation}
is the rotation one form (sometimes also referred to as the extrinsic curvature one form)
with $\Omega_i dx^i \cong \Omega_A dx^A$ the normal fundamental form on $\mathcal{Z}$.

Inserting \eqref{E:xidecomposition} into \eqref{E:qdef}, using \eqref{E:preserving} to evaluate $\beta$ in \eqref{E:qdef}, and also inserting the decomposition \eqref{E:Kappadecomposition} into \eqref{E:qdef},
one finds
\begin{equation}\label{E:current}
	q^i_{\chi}=q_{\chi}\ell^i+X^jS_j{}^i-\theta X^i\,,
\end{equation}
where 
\begin{equation}
	q_{\chi}= \left(\pounds_l f +kf-\theta f)+X^i(\omega_i+\pounds_ln_i\right)\,.
\end{equation}
Taking the  divergence of \eqref{E:current}  we find
\begin{align}
\begin{split}
\label{E:divq}
	\nabla_{i} q_{\chi}^{i} = &
		\pounds_{\ell}\left(\pounds_{\ell} + \kappa\right) f 
		+X^i \pounds_{\ell} \left(\pounds_{\ell} + \kappa\right) n_i + \pounds_X \kappa \\
		&
		+X^i \left(\pounds_{\ell} \omega_i - \partial_i \kappa \right) + \nabla_j \left(X^i \Sigma_{i}{}^{j}\right) \\
		&
		-\theta \left(f (\theta  -  \kappa )- X^i (\omega_i + \pounds_{\ell} n_i )\right)
		- f\pounds_{\ell} \theta  \\
		&
		- \nabla_i \left(\theta X^i\right)\left(1-\frac{1}{d}\right)\,.
\end{split}
\end{align}
Note that the first line on the right hand-side of \eqref{E:divq} vanishes if $\chi^i$ is a generator of a horizon preserving diffeomoprhism, cf. \eqref{E:supers}.

If the event horizon is stationary, that is, there exists an $\alpha$ such that $\tau^a = e^{\alpha} \ell^a$ is a Killing vector near the horizon
\begin{equation}
\label{E:Killing}
	\pounds_{\tau} g_{ab} \Big|_{\mathcal{N}}=0\,,
	\qquad
	\nabla_{c} \pounds_{\tau} g_{ab} \Big|_{\mathcal{N}} = 0\,,
\end{equation}
then
\begin{equation}
	K_{ij} = \pounds_{\ell} g_{ij} = e^{-\alpha} \pounds_{\tau} g_{ij} = 0\,,
\end{equation}
so that $\Sigma_{ij}=0$ and $\theta=0$. In addition, using
\begin{equation}
	\nabla_{a} \nabla_{b} \tau_c \Big|_{\mathcal{N}} = - R_{bca}{}^{d} \tau_d \Big|_{\mathcal{N}}
\end{equation}
(which results from \eqref{E:Killing}), we find that $\pounds_{\tau} \nabla_a \tau^b \Big|_{\mathcal{N}} = 0$ 
implying
\begin{equation}
\label{E:stationarity1}
	\pounds_{e^{\alpha}\ell} \left(\omega_{i} + \nabla_i \alpha\right) = 0\,.
\end{equation}
Further, using the zeroth law of black holes (for stationary horizons) \cite{Bardeen:1973gs}, we have
\begin{equation}
\label{E:stationarity2}
	\partial_i \left(e^{\alpha} \left( \kappa + \pounds_{\ell} \alpha \right) \right) = 0\,.
\end{equation}
Putting together \eqref{E:stationarity1} and \eqref{E:stationarity2} we find
\begin{equation}
	\pounds_{\ell} \omega_i - \partial_i \kappa = 0\,,
\end{equation}
independent on $\alpha$.
Thus, $\nabla_i q_{\chi}^i = 0$ for stationary horizons as long as $\chi$ is a supertranslation or superrotation (or a combination thereof).

\section{Enstrophy and supertranslations}

In an asymptotically AdS geometry, there exist stationary black brane solutions characterized by a uniform Hawking temperature, $T$, and a constant center of mass velocity $u^{\mu}$ relative to an observer at infinity. Here greek indices $\mu,\,\nu,$ denote coordinates on the boundary of AdS space. As discussed in detail in \cite{Bhattacharyya:2008jc}, it is possible to perturb these black brane solutions in a derivative expansion where one assumes that derivatives of $T$ and $u^{\mu}$ are small relative to $T$. The perturbative solution to the Einstein equations takes the form 
\begin{align}
\begin{split}
\label{E:fluidgravitymetric}
	g_{ab}dx^a dx^b =& -r^2 h(r) u_{\mu}u_{\nu} dx^{\mu}dx^{\nu} + r^2 P_{\mu\nu}dx^{\mu}dx^{\nu} \\
		& - 2 u_{\mu}dx^{\mu} dr + \mathcal{O}(\nabla)\,,
\end{split}
\end{align}
where $T$ and $u^{\mu}$ are constrained to satisfy
\begin{equation}
\label{E:EOM}
	\nabla_{\mu} \left(T^d u^{\mu} \right) = \mathcal{O}(\nabla^2)\,,
	\qquad
	Tu^{\nu}\nabla_{\nu}u_{\mu}-P^{\nu}_{\mu}\nabla_{\nu}T={\cal O}(\nabla^2)
\end{equation}
with $\nabla_{\mu}$ a covariant derivative in Minkowski space, $P_{\mu\nu}=\eta_{\mu\nu}+u_{\mu}u_{\nu}$,
and 
\begin{equation}
	h=1-\left(\frac{4\pi T}{(d+1) r}\right)^{d+1}\,.
\end{equation}
Incidentally, \eqref{E:EOM}  correspond to the hydrodynamic equations \eqref{E:TEOM} (see \cite{Bhattacharyya:2008jc}). 

The location of the event horizon for the geometry \eqref{E:fluidgravitymetric} is given by 
\begin{equation}
	r = \frac{4\pi T}{d+1} + \mathcal{O}(\nabla)\,.
\end{equation}
Here and in the remainder of this section we will use the coordinates $x^{\mu}$ to parameterize the horizon.
It now follows that
\begin{equation}
\label{E:defell}
	\ell^i \partial_i =  u^{\mu}\partial_{\mu} + \mathcal{O}(\nabla)
\end{equation}
and also
\begin{equation}
	\kappa = 2 \pi T +\mathcal{O}(\nabla)\,.
\end{equation}
 We will also choose,
\begin{equation}
\label{E:defn}
	n_i dx^i =   u_{\mu} dx^{\mu} + \mathcal{O}(\nabla)\,.
\end{equation}
(Note that this choice of $n_i$ satisfies \eqref{E:n} to leading order in a derivative expansion.)
With this parameterization we find that $\Omega_A = \mathcal{O}(\nabla)$ and that
\begin{equation}
	\Sigma_i{}^j dx^i \partial_j = {\frac{1}{2}}\sigma_{\mu}{}^{\nu} dx^{\mu} \partial_{\nu}+{\cal O}(\nabla^2)\,,
\end{equation}
where $\sigma_{\mu\nu}$ 
is the shear tensor for a fluid with velocity $u^{\mu}$,
\begin{equation}
\label{E:defshear}
	\sigma^{\mu\nu} = P^{\mu\alpha}P^{\nu\beta}\left(\nabla_{\alpha}u_{\beta} + \nabla_{\beta}u_{\alpha}\right) - \frac{2}{d}P^{\mu\nu} \nabla_{\alpha}u^{\alpha}\,.
\end{equation}
With our choice of parameterization we find that the Einstein equations take the form
\begin{equation}
\label{E:EOMscaling}
	\theta = \mathcal{O}(\nabla^2)\,,
	\qquad
	\left(\pounds_{\ell}\omega_i - \partial_i \kappa\right)\Pi^i_A = \mathcal{O}(\nabla^2)\,.
\end{equation}

Suppose that $\chi^i$ is a supertranslation ($X^i=0$) and that $f$ is order $\mathcal{O}(\nabla^{[f]})$ in a derivative expansion. In this case $q_{\chi}^i$ is also $\mathcal{O}(\nabla^{[f]})$ but 
\begin{equation}
\label{E:divqsupert}
	\nabla_i q_{f u}^i = \mathcal{O}\left(\nabla^{[f]+2}\right)
\end{equation}
on account of the equation of motion \eqref{E:EOMscaling}. Thus, supertranslations lead to an approximately conserved current in the derivative expansion. 
Superrotations will not lead to approximately conserved currents owing to the fact that $\nabla_i q_X^i = \mathcal{O}(\nabla^{[X]+2})$ but $q_X^i = \mathcal{O}(\nabla^{[X]+1})$.

In a derivative expansion the supertranslation constraint, \eqref{E:supers}, reads
\begin{equation}\label{E:supertranslationconstraint}
	\pounds_{u} (\kappa f) = \mathcal{O}(\nabla^{[f]+2})
\end{equation}
and the resulting current is
\begin{equation}
\label{E:truncatedsuper}
	q_{f u}^i \partial_i =  \kappa f u^{\mu}\partial_{\mu}\,.
\end{equation}
The divergence of $q_{f u}^i$ is given, in our current coordinate system, by
\begin{equation}
\label{E:divsuper}
	\nabla_i q^i_{fu} = \frac{1}{T^{d}} \partial_i \left( T^{d} q^i_{fu} \right)\,,
\end{equation}
where the factors of $T^{d}$ come from the measure on the spatial section of the horizon. We are guaranteed that $q^i_{fu}$ is approximately conserved  provided that \eqref{E:supertranslationconstraint} is satisfied which implies that the divergence of
\begin{equation}
\label{E:Jfromgravity}
	J_f^{\mu} = T^{d} \kappa f u^{\mu}
\end{equation}
in Minkowski space will approximately vanish.

Equation \eqref{E:Jfromgravity} and its approximate conservation reproduces \eqref{E:divJf} once we identify $J^{\mu}_f$ in \eqref{E:Jfromgravity} with $J_g^{\mu}$ in \eqref{E:Jgdef}.
Thus, approximate enstrophy conservation  is a result of a particular set of supertranslations on the horizon of AdS${}_4$ black branes 
 which are local in the black hole temperature and center of mass velocity, $T$ and $u^{\mu}$. Likewise, approximate entropy conservation is a result of horizon supertranslations of AdS${}_{d+2}$ black branes with $f=c/\kappa$, (with $c$ a constant).

\section{Discussion and outlook}

In this work we have shown that horizon supertranslation generators of AdS black brane geometries, which are conserved in the stationary limit, are associated with approximately conserved  currents in a dual fluid  description of the geometry.

In asymptotically AdS${}_4$ geometries there exist a small subset of supertranslation generators which are local functions of the Hawking temperature and black brane null generators which are dual to the approximately conserved enstrophy and its various moments. Horizon superrotations are not associated with conserved charges of the dual fluid since  conservation of superrotations in the stationary limit is not enhanced to an approximate symmetry once stationarity is only approximate.

The analysis carried out in this work pertains to asymptotically AdS black branes in the absence of matter, dual to uncharged conformal fluids. We expect our main result to be applicable more broadly to charged non conformal fluids. More generally, relying on the membrane paradigm \cite{damour1979quelques}, one might expect a similar construction for black holes in asymptotically flat space, or even black holes in general.  

Non relativistic enstrophy is not only conserved in the absence of dissipation, it can not increase in time. Whether the same can be said regarding the relativistic enstrophy current is yet an open problem. Another hydrodynamic quantity which we know must not decrease in time (in a relativistic setting or not) is the entropy. That entropy can not decrease translates into the well known area increase theorem of black holes \cite{Bekenstein:1972tm}. Thus, one can not help but wonder whether there is an enstrophy decrease theorem for black holes valid, at the very least, at low velocities. While the present work has not dealt with a full description of black hole enstrophy and its dynamics, we hope it will provide a stepping stone towards it.

\section*{Acknowledgments}

We would like to thank A. Frishman for useful discussions. 	
NPF is supported by the European
Commission through the Marie Sklodowska-Curie Action UniCHydro (grant agreement ID: 886540). NPF would also like to acknowledge support from the Center for Theoretical Physics and the Department of Physics at the Massachusetts Institute of Technology. AY and BT are supported in part by an Israeli Science Foundation excellence center grant 2289/18 and a Binational Science Foundation grant 2016324.

\bibliography{EShort}

\end{document}